\documentclass[12pt]{article}
\usepackage{hyperref}
\usepackage{amsmath,amssymb,amsfonts,amsthm,amscd,textcomp}
\usepackage{longtable}
\usepackage[top=15mm,bottom=15mm,left=17.5mm,right=17.5mm]{geometry}
\usepackage{flushend}
\usepackage{multicol}
\usepackage{cite}
\usepackage{longtable}
\usepackage{threeparttable}
\usepackage[dvips]{color}
\usepackage{color}
\usepackage{setspace}
\usepackage{balance}
\usepackage{lastpage}
\usepackage{graphicx}
\usepackage{subfigure}
\usepackage{caption}
\usepackage{nomencl}
\usepackage{pstricks}
\usepackage{pstricks-add}
\usepackage{pst-plot,pstricks-add}
\usepackage{multirow}
\usepackage{rotating}
\title{Comparative analysis of two new wind speed T-X models using Weibull and log-logistic distributions for wind energy potential estimation in Tabriz, Iran}
\author{{Meysam Mohammadpour and Hossein Bevrani}\footnote{Corresponding author,   Email:bevrani@gmail.com}\\
\centering{\small{Faculty of Mathematics, Statistics and Computer Sciences, University of Tabriz, Tabriz, Iran}}}
\date{}
\begin{document}
\maketitle
\begin{abstract}
To assess the potential of wind energy in a specific area, statistical distribution functions are commonly used to characterize wind speed distributions. The selection of an appropriate wind speed model is crucial in minimizing wind power estimation errors. In this paper, we propose a novel method that utilizes the T-X family of continuous distributions to generate two new wind speed distribution functions, which have not been previously explored in the wind energy literature. These two statistical distributions, namely the Weibull-three parameters-log-logistic (WE3-LL3) and log-logistic-three parameters-Weibull (LL3-WE3), are compared with four other probability density functions (PDFs) to analyze wind speed data collected in Tabriz, Iran. The parameters of the considered distributions are estimated using maximum likelihood estimators with the Nelder-Mead numerical method. The suitability of the proposed distributions for the actual wind speed data is evaluated based on criteria such as root mean square errors, coefficient of determination, Kolmogorov-Smirnov test, and chi-square test. The analysis results indicate that the LL3-WE3 distribution demonstrates generally superior performance in capturing seasonal and annual wind speed data, except for summer, while the WE3-LL3 distribution exhibits the best fit for summer. It is also observed that both the LL3-WE3 and WE3-LL3 distributions effectively describe wind speed data in terms of the wind power density error criterion. Overall, the LL3-WE3 and WE3-LL3 models offer a highly accurate fit compared to other PDFs for estimating wind energy potential.
\end{abstract}
\section{Introduction} 
The use of energy is continuously expanding due to the growing global population and technological developments. In recent years, increasing the usage of fossil fuels in most the fields of life such as industry, heating, and transportation led to reserves of fossil fuels gradually reduced and are not always available. Further, they create detriments in terms of the environment, including climate change, carbon emission, and air pollution. Also obtaining and the use of them are expensive \cite{ref1,ref2}. These drawbacks show that countries should be to seek renewable and clean energy sources (such as hydro, geothermal, solar, biomass, and wind energy), which are economical and environmentally friendly. Unlike fossil fuels, these energies remove the risk of increase in costs, decreasing carbon emissions, and also many of them are everlasting \cite{ref1,ref2,ref3,ref4}.\\
In this research, from among various eco-friendly renewable energy sources, wind energy is considered. Wind energy is an inexpensive and unlimited natural resource and it is not noxious for the environment as compared to fossil fuels. Besides, the usage of wind power does not require complex technology. The development and investment in wind energy will provide advantages in terms of economic activities, opportunities for employment, energy independence in the electric power generation sector, studies and research, and easy transportation \cite{ref2,ref3,ref4}. Nowadays, rapid research and developments in the field of wind turbine technology have caused significant growth in electricity production across the world \cite{ref1,ref4}. So for these reasons, wind energy applies as a major resource of renewable energies in approximately 103 countries in the world \cite{ref5,ref6}.\\
By using two important factors, wind energy can be obtained economically and efficiently. The first factor is the selection of a suitable windy location for the installation of the turbines and the second factor is to determine the specifications of wind speed via the best distribution functions \cite{ref1,ref5,ref7,ref8}. The choice of the appropriate distribution of wind speed is a necessary stage for the calculation of wind energy potential because the generation of wind power is acutely dependent on the capacity and the characteristics of wind magnitude \cite{ref2,ref4,ref5}. In wind energy studies, the most common and the most popular distribution function for modeling wind speed data is 2-parameter Weibull distribution \cite{ref1,ref9,ref10,ref11,ref12,ref13,ref14,ref15,ref16}. Estimated parameters of Weibull distribution are yielded with the various methods, such as the least squares estimators (LSEs), maximum likelihood estimators (MLEs), and method of moments (MM) see for example \cite{ref7,ref10,ref17,ref18,ref19}.\\
Notwithstanding extensive usage of Rayleigh and 2-parameters Weibull distributions in most studies, they cannot always be suitable for modeling wind speed data in all windy regimes in nature \cite{ref1,ref5}. Based on other researchers, in order to the wind energy estimation, various statistical models have been applied to determine wind speed probability distributions. \cite{ref20,ref21,ref22,ref23,ref24,ref25,ref26,ref27,ref28,ref29}.These distributions are the Pearson type III (P3), Wakaby (WA), Gamma (G), 2 and 3-parameter Lognormal (LN2, LN3), Inverse Gamma (IG), Log-Pearson type III (LP3), Inverse Gaussian (IGA), 3-parameter Beta (B), Erlang (ER), Kappa (KAP), Burr (BR), and Gumbel (EV1). Therefore, recognizing the suitable probabilistic distributions that have a better performance to the wind data is of essential matter.\\
One of the issues presented together with the chosen wind speed model is the parameter estimation method for distributional models \cite{ref28}. Accurate calculation of distributional parameters is essential because measured wind energy potential depends on the estimated parameters of the selected PDFs. Some estimation methods have been applied and evaluated in terms of various criteria. Among these methods, when the sample size is greater than 100, therefore the maximum likelihood estimation (MLE) method is generally suitable. As a result, we choose MLE to estimate the parameters of the models.\\
In recent decades, many popular distributions are applied for modeling the wind speed data set, but they cannot always explain enough flexibility to show an appropriate fit in various geographical locations of the world. For this reason, new methods are presented and employed to obtain generalizations of well-known distributions. In this research, we introduce a new approach to generate two new statistical distributions by using the T-X family of continuous distributions that are proposed for the first time to assess the wind speed distribution in the wind energy literature. Two new PDFs will be produced by combining 3-parameter Weibull and 3-parameter log-logistic distribution functions. These distributions are Weibull-three parameters-log-logistic (WE3-LL3) and log-logistic-three parameters-Weibulll (LL3-WE3).\\
The main idea of this study is to the comparison of the effectiveness of the generated distributions versus some of the most common distribution functions namely 3-parameter Weibull (WE3), 3-parameter Log-Normal (LN3), 3-parameter log-logistic (LL3), and Generalized Extreme Value (GEV) based on model selection criteria, such as  the root mean square errors (RMSE), coefficient of determination ($R^2$), Kolmogorov-Smirnov (KS) test, and chi-square ($\chi^2$) test for modeling wind speed data measured at Tabriz in Iran. In other words, we try to identify the most proper probability density functions that illustrate very good performance for wind speed data analysis.\\
The remainder of the present paper is structured as follows: Section \ref{sec2} presents the analysis of the recorded wind speed data. Mathematical models applied to describe the wind speed probability function are proposed in Section \ref{sec3}. Section \ref{sec4} provides the modeling methodology of this study, including the distribution parameters estimation and the model selection criteria. The results of the obtained evaluations using the R programming software are discussed in Section \ref{sec5}. Finally, Section \ref{sec6} includes comments and conclusions.     
\section{Wind speed data}\label{sec2}
Tabriz is situated in the northwest of Iran, in the East Azerbaijan province. It is positioned between the Eynali and Sahand mountains in a fertile region, along the shores of the Ghuri River and Aji River. The elevation of Tabriz varies from 1,350 to 1,600 meters (4,430 to 5,250 ft) above sea level, and the city covers an area of approximately 240 km$^2$. Tabriz experiences a humid continental climate, characterized by distinct seasons. Summers are dry and semi-hot, winters are snowy and cold, autumns are rainy and humid, and springs have mild weather. The average annual temperature in Tabriz is approximately 12.6 °C (54.7 °F).\\ 
In the present research, the wind speed data was recorded in three hours time intervals at 10 m height in the 2018 year from Tabriz station with the latitude 38°08\textasciiacute02\textacutedbl N, the longitude 46°14\textasciiacute06\textacutedbl E and the elevation 1.359 (m), is taken from Iran State Meteorological Service. In the wind speed data set, instead of missing values, the known observations data was replaced. This method is simple and it can provide a better result for missing data with random behaviors \cite{ref30}. The descriptive statistics such as maximum, mean, standard deviation (SD), standard error of the mean(SE.Mean), kurtosis, skewness, and quarters ($Q_{1}$,$Q_{2}$,$Q_{3}$) as brief to identify the windy characteristics are demonstrated in table \ref{Tab1}.   
\begin{table}[h!]
\begin{center}\caption{\footnotesize{Values of statistical quantities for the reported wind speed data (m/s) from Tabriz station.}}
\renewcommand{\arraystretch}{1.1}
\resizebox{\textwidth}{!}{
\begin{tabular}{cccccccccccc}
\hline
Year&Season&n&Max (m/s)&Mean (m/s)&SD (m/s)&SE.Mean&Skewness&Kurtosis& $Q_{1}$&$Q_{2}$&$Q_{3}$\\
\hline
2018&&2920&18.07&3.63&2.38&0.04&1.16&4.73&1.99&3.01&4.99\\
&Winter&720&16.77&2.63&1.69&0.06&2.29&14.23&1.93&2.12&3.04\\
&Spring&736&18.07&3.86&2.55&0.09&1.36&5.45&2.02&3.04&5.00\\
&Summer&736&13.00&4.73&2.41&0.09&0.62&3.00&2.96&4.09&6.10\\
&Autumn&728&11.95&3.29&2.25&0.08&0.90&3.62&1.94&2.94&4.86\\
\hline
\end{tabular}}\label{Tab1}
\end{center}
\end{table}
\section{Mathematical models}\label{sec3}
\subsection{One-component probability distributions}
Usually, having science on the wind speed density function is of essential importance to evaluate the wind energy potential, efficiency of wind energy systems, and describe the wind speed characteristics. Thus, it is very significant to select the most favorable models that show better modeling performance to the wind speed data. In this subsection, we introduce four distribution functions that are including WE3, LN3, LL3, and GEV distributions.\\ 
The one-component statistical models suggested in this paper are reviewed here.
\subsubsection{Three parameters Weibull distribution}
The WE3 distribution is a continuous probability distribution with an additional location parameter\cite{ref23}. The PDF of the WE3 is given below:
\begin{equation}
f_{WE3}(x,\mu,\omega,\delta)=\frac{\omega}{\mu}\left[\frac{x-\delta}{\mu}\right]^{\omega-1}\exp\left[-\left(\frac{x-\delta}{\mu}\right)^{\omega}\right] \,; \quad x>\delta \, , \ \mu,\omega>0
\end{equation} 
The CDF of WE3 is obtained by
\begin{equation}\label{eq00}
F_{WE3}(x)=1-\exp\left[-\left(\frac{x-\delta}{\mu}\right)^{\omega}\right]
\end{equation}
where $x$ is the wind speed, $\mu$, $\omega$, and $\delta$ are the shape, scale, and location parameters, respectively.
\subsubsection{Three parameters log-logistic distribution}
suppose Y has a logistic distribution with the shape and scale parameters, so X=$\exp(\mathrm{Y})+\delta$ has a LL3 distribution with the shape and scale parameters and an additional location parameter $\delta$ \cite{ref33}. The LL3 PDF is described as follows:
\begin{equation}
f_{LL3}(x,\mu,\omega,\delta)=\frac{\left(\frac{\omega}{\mu}\right)\left(\frac{x-\delta}{\mu}\right)^{\omega-1}}{\left(1+\left(\frac{x-\delta}{\mu}\right)^\omega\right)^2}\,; \quad x>\delta \, ,\ \mu>0\, ,\ \omega\ge 1
\end{equation}
Its CDF is expressed as
\begin{equation}\label{eq01}
F_{LL3}(x)=\frac{1}{1+\left(\frac{x-\delta}{\mu}\right)^{-\omega}}=\frac{\left(\frac{x-\delta}{\mu}\right)^{\omega}}{1+\left(\frac{x-\delta}{\mu}\right)^{\omega}}
\end{equation}
where $x$ is the wind speed, $\mu$, $\omega$, and $\delta$ are the shape, scale, and location parameters, respectively.
\subsubsection{Three parameters log-normal distribution}
If random variable Y is distributed according to a normal distribution with the shape (mean) and scale (standard deviation) parameters, then X=$\exp(\mathrm{Y})+\delta$ has a LN3 distribution with the shape and scale parameters and an additional location parameter $\delta$ \cite{ref34}. The PDF of LN3 is defined as
\begin{equation}
f_{LN3}(x,\mu,\omega,\delta)=\frac{1}{(x-\delta)\omega\sqrt{2\pi}}\exp\left[-\frac{\left(\ln(x-\delta)-\mu\right)^{2}}{2\omega^2}\right] \,; \quad x>\delta \, , \; \mu \in \mathbb{R} \,,\ \omega>0
\end{equation} 
Also, the LN3 CDF is written as
\begin{equation}
F_{LN3}(x)=\int_{\delta}^{x} \frac{1}{(t-\delta)\omega\sqrt{2\pi}}\exp\left[-\frac{\left(\ln(t-\delta)-\mu\right)^{2}}{2\omega^2}\right]dt
\end{equation}
where $x$ is the wind speed, $\mu$, $\omega$, and $\delta$ are the shape, scale, and location parameters, respectively.
\subsubsection{Generalized Extreme Value distribution}
The GEV distribution is a continuous probability distribution and flexible that combines the Weibull, Fréchet, and Gumbel models \cite{ref35}. The PDF of the GEV is obtained by
\begin{equation}
f_{GEV}(x,\mu,\omega,\delta)=\frac{1}{\omega}\left[1+\mu\left(\frac{x-\delta}{\omega}\right)\right]^{-\frac{1}{\mu}-1}\exp\left[-\left[1+\mu\left(\frac{x-\delta}{\omega}\right)\right]^{-\frac{1}{\mu}}\right]
\end{equation}
And the GEV CDF is given below:
\begin{equation}
F_{GEV}(x)=\exp\left[-\left[1+\mu\left(\frac{x-\delta}{\omega}\right)\right]^{-\frac{1}{\mu}}\right]
\end{equation}
where $x$ is the wind speed, $\mu \in \mathbb{R}$, $\omega>0$, and $\delta \in \mathbb{R}$ are the shape, scale, and location parameters, respectively. It should be noted when $\mu>0$ then $x\in [\delta-\frac{\omega}{\mu},\infty)$ and when $\mu<0$ then $x\in (-\infty,\delta-\frac{\omega}{\mu}]$.
\subsection{Family of T-X continuous distributions}
Mathematical distribution functions are usually employed  to characterize real world phenomena. Because of the efficiency of statistical models, their theory is widely studied and some popular methods have been extended for generating a new class of distributions. Here first, we present the definition of the T-X family of continuous distributions and then, will generate two new distributions of WE3 and LL3 models to evaluate the wind speed data.\\
Suppose $h(x)$ and $H(x)$ are the probability density function and cumulative distribution function of the random variable X, respectively. Also, let $k(t)$ be the PDF of a continuous random variable $T \in [m,n]$, for $-\infty \le m<n\le \infty$. The cumulative distribution function of the new T-X family of distributions is expressed using the below function \cite{ref31}:
\begin{equation}\label{eq0}
F(x)=\int_{m}^{G(H(x))}k(t)dt=K(t)=K(G(H(x)))
\end{equation}
where T has CDF $K(t)$, $t=G(H(x))$ is a function of the CDF $H(x)$, and it should satisfy in the following conditions\cite{ref31}:
\begin{equation*}
\begin{split}
&G(H(x)) \in [m,n]\\
&G(H(x)) \quad \text{is differentiable and monotonically non-decreasing}\\
&G(H(x)) 
\begin{CD} 
@>>> m \ \, \text{as} \ \, x @>>> -\infty 
\end{CD} \,\
\text{and} \,\
G(H(x))
\begin{CD} 
@>>> n \ \, \text{as} \ \, x @>>> \infty 
\end{CD}
\end{split}
\end{equation*}
The PDF of $F(x)$ is formulated as
\begin{equation}\label{eq1}
f(x)=\left[\frac{d}{dx}G(H(x))\right]k(G(H(x)))
\end{equation}
It should be noted that $F(x)$ is a composite function of $K(t)=K(G(H(x)))$ and also the probability density function $k(t)$ in equation \eqref{eq0} is transformed into a new cumulative distribution function $F(x)$ via the function $G(H(x))$, which acts as a transformer. Therefore, distribution $f(x)$ in \eqref{eq1} is transformed from random variable T by using the transformer random variable X and call it T-X or Transformed-Transformer distribution \cite{ref32}.\\
Various $G(H(x))$ functions present different families of T-X distributions. The definition of $G(H(x))$ depends on the support of the random variable T. In this study, we consider the support of T as $[m,\infty)$, $m\ge0$. Then, $G(H(x))$ is defined by the below function \cite{ref31}: 
\[G(H(x))=\frac{H(x)}{1-H(x)}\]
With replacing $G(H(x))$ in equation \eqref{eq1}, $f(x)$ is as follows:
\begin{equation}\label{eq11}
f(x)=\left[\frac{h(x)}{\left(1-H(x)\right)^2}\right]k\left(\frac{H(x)}{1-H(x)}\right)
\end{equation}
Now, we introduce two new distributions that are obtained through the WE3 and LL3 models in the following subsections.
\subsubsection{\small Weibull-three parameters-log-logistic distribution}
\begin{figure}[htp!]
	\begin{center}
		\subfigure{
			\includegraphics[height=6.5cm,width=7cm]{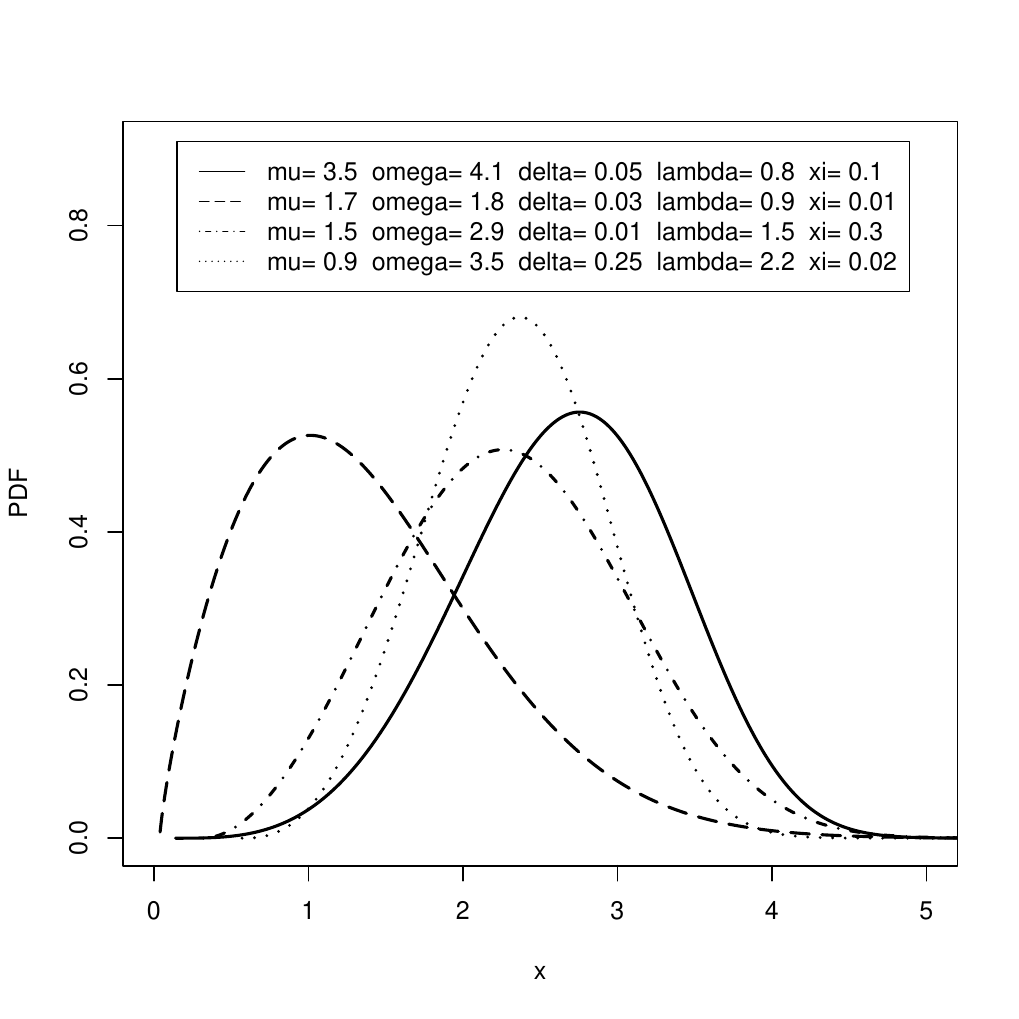}}
		\hspace{0.5cm}
		\subfigure{
			\includegraphics[height=6.5cm,width=7cm]{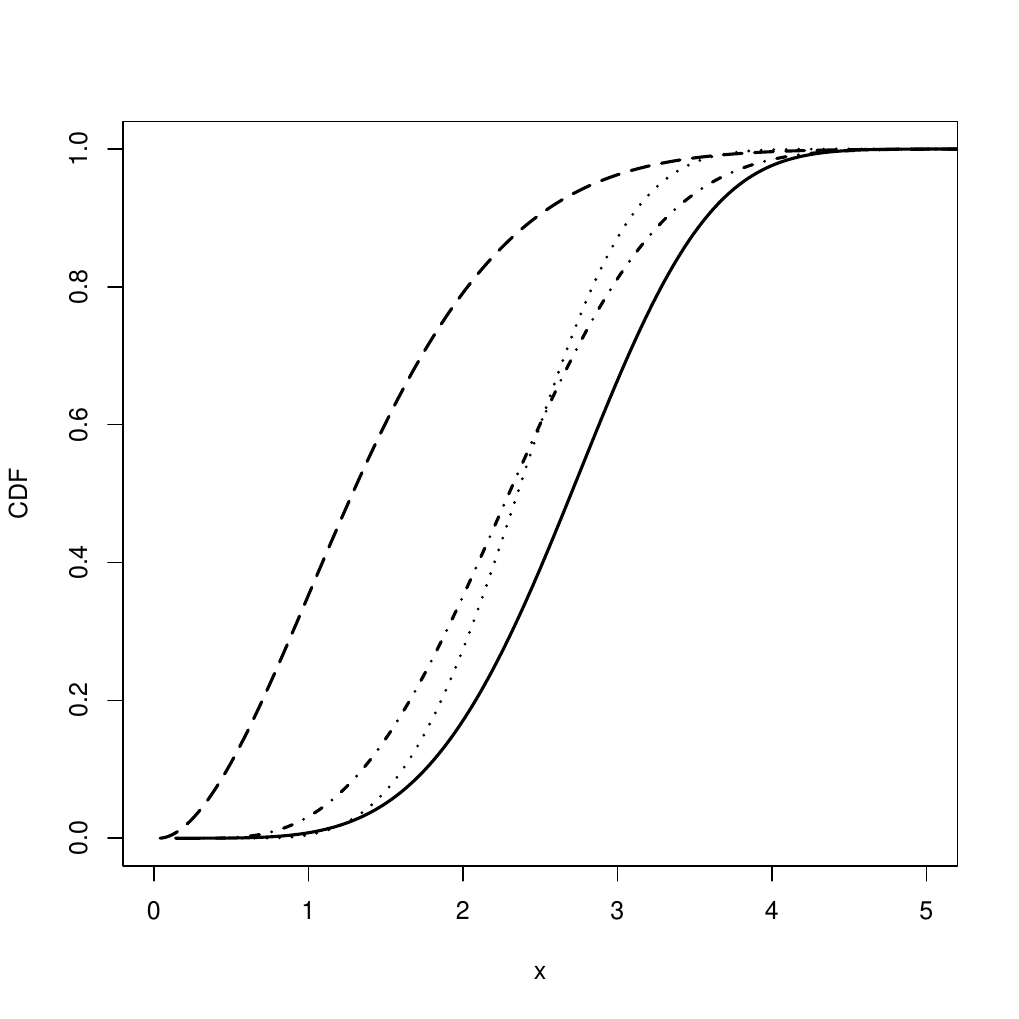}}
		\hspace{0.5cm}	
	\end{center}\caption{\small
	The CDF and the PDF plots of the WE3-LL3 distribution.}
\label{fig0}
\end{figure}

suppose X has three parameters log-logistic distribution with PDF $h(x)$ and CDF $H(x)$ with $\lambda$, $\beta$, and $\xi$ parameters and also if random variable T be distributed according to Three parameters Weibull distribution with PDF $k(t)$ and CDF $K(t)$ with $\mu$, $\omega$, and $\delta$  parameters. So the CDF of WE3-LL3 is formulated as
\begin{equation}
F(x)=\int_{\delta}^{\frac{H(x)}{1-H(x)}}k(t)dt=K\left(\frac{H(x)}{1-H(x)}\right)-K(\delta)
\end{equation}
With replacing $\delta$ in equation \eqref{eq00}, we will have $K(\delta)=0$, so $F(x)$ is as follows:
\begin{equation}
\begin{split}
F(x)&=\int_{\delta}^{\frac{H(x)}{1-H(x)}}k(t)dt=K\left(\frac{H(x)}{1-H(x)}\right)=1-\exp\left[-\left(\frac{\frac{H(x)}{1-H(x)}-\delta}{\mu}\right)^\omega\right]\\
&=1-\exp\left[-\left(\frac{(1+\delta)H(x)-\delta}{\mu(1-H(x))}\right)^\omega\right]\overset{H_{LL3}(x)}{=\joinrel=\joinrel=\joinrel=}1-\exp\left[-\left(\frac{(x-\xi)^{\beta}-\delta\lambda^\beta}{\mu\lambda^\beta}\right)^\beta\right]
\end{split}
\end{equation}
The corresponding PDF of WE3-LL3 distribution with respect to equation \eqref{eq11} is as
\begin{equation}
f(x,\mu,\omega,\delta,\lambda,\beta,\xi)=\left(\frac{\omega}{\mu^\omega}\right)\frac{\beta(x-\xi)^{\beta-1}}{\lambda^\beta}\left(\frac{(x-\xi)^{\beta}-\delta\lambda^\beta}{\lambda^\beta}\right)^{\omega-1} \exp\left[-\left(\frac{(x-\xi)^{\beta}-\delta\lambda^\beta}{\mu\lambda^\beta}\right)^\omega\right]
\end{equation}
Where $x$ is the wind speed, $\mu,\omega,\lambda>0$, $\beta\ge1$, $\delta\in \mathbb{R}$, and $x\ge\xi$. The CDF and the PDF plots of the WE3-LL3 distribution are shown in Fig. \ref{fig0} and are obtained for various values of the $\mu,\omega,\delta,\lambda$, and $\xi$ parameters and $\beta=1$. Also, the WE3-LL3 distribution has the following quantile function:
\begin{equation}
Q(p|\mu,\omega,\delta,\lambda,\beta,\xi)=\lambda\left[\mu\left(-\log(1-p)\right)^{\frac{1}{\omega}}+\delta\right]^{\frac{1}{\beta}}+\xi \quad 0\le p \le 1
\end{equation}
\subsubsection{\small Log-logistic-three parameters-Weibull distribution}
\begin{figure}[htp!]
	\begin{center}
		\subfigure{
			\includegraphics[height=6.5cm,width=7cm]{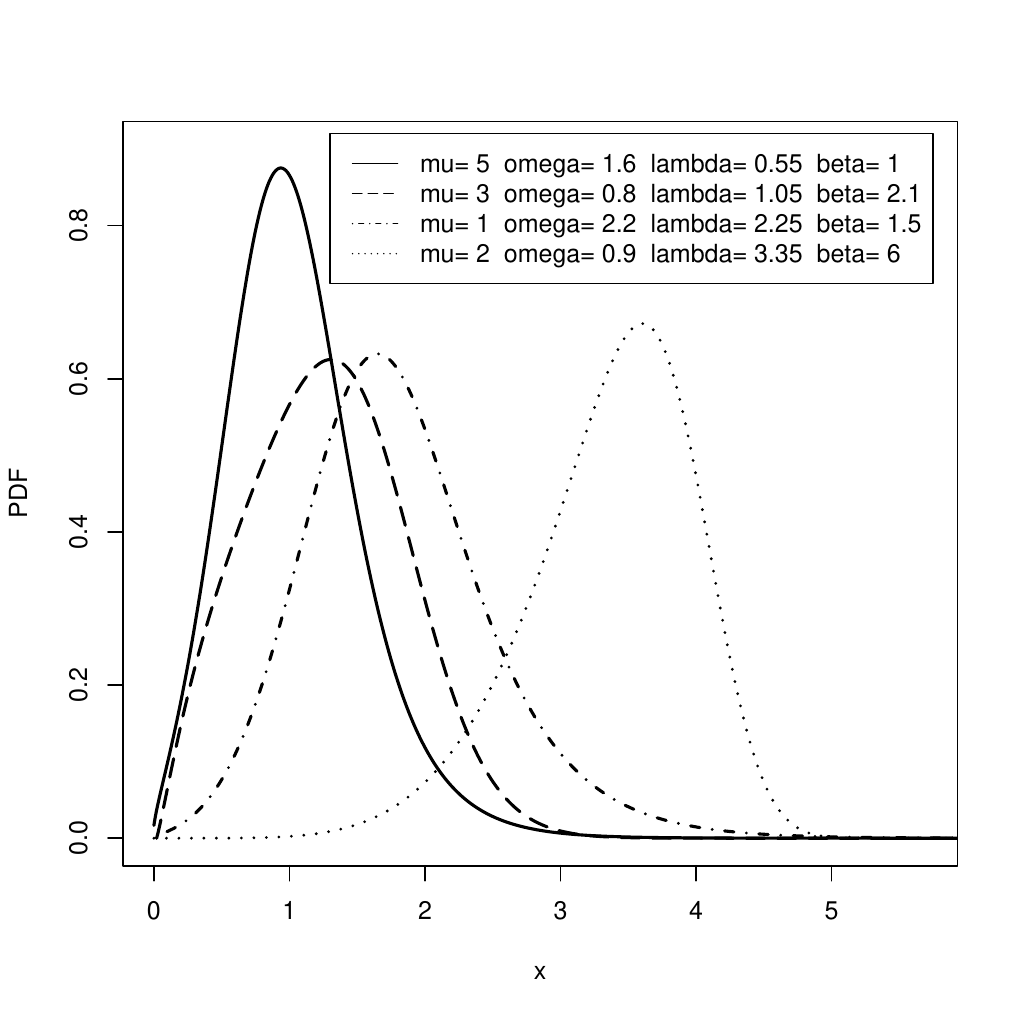}}
		\hspace{0.5cm}
		\subfigure{
			\includegraphics[height=6.5cm,width=7cm]{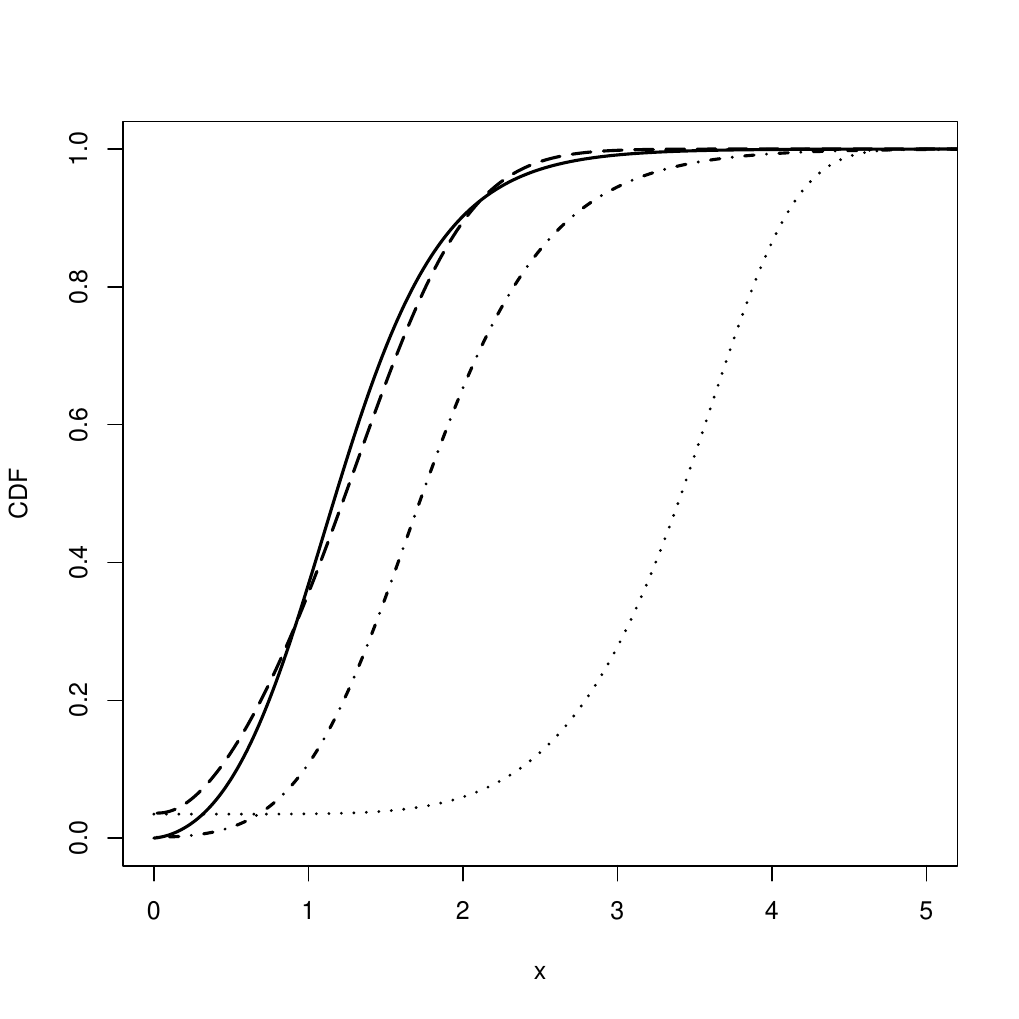}}
		\hspace{0.5cm}	
	\end{center}\caption{\small
	The CDF and the PDF plots of the LL3-WE3 distribution.}
\label{fig1}
\end{figure}

suppose random variable X is distributed according to Three parameters Weibull distribution with PDF $h(x)$ and CDF $H(x)$ with $\lambda$, $\beta$, and $\xi$ parameters and also suppose T has three parameters log-logistic distribution with PDF $k(t)$ and CDF $K(t)$ with $\mu$, $\omega$, and $\delta$ parameters. Then, the CDF of the LL3-WE3 distribution is defined as
\begin{equation}
F(x)=\int_{\delta}^{\frac{H(x)}{1-H(x)}}k(t)dt=K\left(\frac{H(x)}{1-H(x)}\right)-K(\delta)
\end{equation}
With replacing $\delta$ in equation \eqref{eq01}, we will have $K(\delta)=0$, therefore $F(x)$ is as
\begin{equation}
\begin{split}
F(x)=\int_{\delta}^{\frac{H(x)}{1-H(x)}}k(t)dt&=K\left(\frac{H(x)}{1-H(x)}\right)=\frac{\left(\frac{H(x)}{1-H(x)}-\delta\right)^\omega}{\mu^\omega+\left(\frac{H(x)}{1-H(x)}-\delta\right)^\omega}=\frac{1}{1+\left(\frac{\mu(1-H(x))}{(1+\delta)H(x)-\delta}\right)^\omega}\\
&\overset{H_{WE3}(x)}{=\joinrel=\joinrel=\joinrel=\joinrel=}\frac{1}{1+\left(\frac{\mu\exp\left[-\left(\frac{x-\xi}{\lambda}\right)^{\beta}\right]}{1-(1+\delta)\exp\left[-\left(\frac{x-\xi}{\lambda}\right)^{\beta}\right]}\right)^\omega}=\frac{1}{1+\left(\frac{\mu}{\frac{1}{\exp\left[-\left(\frac{x-\xi}{\lambda}\right)^{\beta}\right]}-(1+\delta)}\right)^\omega}
\end{split}
\end{equation}
Also, the PDF of the LL3-WE3 with respect to equation \eqref{eq11} is expressed by the below function:
\begin{equation}
f(x,\mu,\omega,\delta,\lambda,\beta,\xi)=\omega\left(\frac{\beta}{\lambda}\right)\frac{\left(\frac{x-\xi}{\lambda}\right)^{\beta-1}\left(\mu\exp\left[-\left(\frac{x-\xi}{\lambda}\right)^{\beta}\right]\right)^{\omega}\left(1-(1+\delta)\exp\left[-\left(\frac{x-\xi}{\lambda}\right)^{\beta}\right]\right)^{\omega-1}}{\left[\left(1-(1+\delta)\exp\left[-\left(\frac{x-\xi}{\lambda}\right)^{\beta}\right]\right)^{\omega}+\left(\mu\exp\left[-\left(\frac{x-\xi}{\lambda}\right)^{\beta}\right]\right)^{\omega}\right]^2}
\end{equation}
Where $x$ is the wind speed, $\mu,\lambda,\beta>0$, $\omega\ge1$, $\delta\in \mathbb{R}$, and $x\ge\xi$. The CDF and the PDF plots of the LL3-WE3 distribution are demonstrated in Fig. \ref{fig1} and are obtained for various values of the $\mu,\omega,\lambda$, and $\beta$ parameters and $\delta=-0.05, \xi=0.02$. Also, the LL3-WE3 distribution has the quantile function as follows:
\begin{equation}
Q(p|\mu,\omega,\delta,\lambda,\beta,\xi)=\lambda\left[-\log\left(\frac{(1-p)^\frac{1}{\omega}}{(1+\delta)(1-p)^\frac{1}{\omega}+\mu p^\frac{1}{\omega}}\right)\right]^{\frac{1}{\beta}}+\xi \quad 0\le p \le 1
\end{equation}
\section{Methodology}\label{sec4}
\subsection{Parameters estimation}
In this subsection, is presented the summary description of the MLEs to estimate the unknown parameters of the WE3, LL3, LN3, GEV, WE3-LL3, and LL3-WE3 distributions. In the MLEs method, It is clear that the ML estimates of the parameters are determined as the point that the log-likelihood function maximizes for the considered point \cite{ref1}. Thus, we provide the log-likelihood functions for the suggested distributions for the random sample $X_{1},X_{2},\dots,X_{n}$. The
log-likelihood functions are formulated as
\begin{equation}\label{eq3}
\ln{L}_{WE3}=n \ln(\omega)-n\omega\ln(\mu)+(\omega-1)\sum_{i=1}^{n}\ln(x_{i}-\delta)-\mu^{-\omega}\sum_{i=1}^{n}(x_{i}-\delta)^{\omega}
\end{equation}
\begin{equation}\label{eq4}
\ln{L}_{LL3}=n\ln(\omega)-n\omega\ln(\mu)+(\omega-1)\sum_{i=1}^{n}\ln\left(x_{i}-\delta\right)-2\sum_{i=1}^{n}\ln\left(1+\left(\frac{x_{i}-\delta}{\mu}\right)^{\omega}\right)
\end{equation}
\begin{equation}\label{eq5}
\ln{L}_{LN3}=-\frac{n}{2}\ln(2\pi)-n\ln(\omega)-\sum_{i=1}^{n}\ln(x_{i}-\delta)-\frac{\sum_{i=1}^{n}\left(\ln(x_{i}-\delta)-\mu\right)^2}{2\omega^2}
\end{equation}
\begin{equation}\label{eq6}
\ln{L}_{GEV}=-n\ln(\omega)+\sum_{i=1}^{n}\left[\left(-\frac{1}{\mu}-1\right)\ln\left(1+\mu\left[\frac{x_{i}-\delta}{\omega}\right]\right)-\left[1+\mu\left(\frac{x_{i}-\delta}{\omega}\right)\right]^{-\frac{1}{\mu}}\right]
\end{equation}
\begin{equation}\label{eq7}
\begin{split}
\ln{L}_{WE3-LL3}&=n\ln\omega-n\omega\ln\mu+n\ln\beta+(\beta-1)\sum_{i=1}^{n}\ln(x_{i}-\xi)-n\omega\beta\ln\lambda\\
&+(\omega-1)\sum_{i=1}^{n}\ln\left(\left[(x_{i}-\xi)^{\beta}-\delta\lambda^\beta\right]\right)-\sum_{i=1}^{n}\left(\frac{(x_{i}-\xi)^{\beta}-\delta\lambda^\beta}{\mu\lambda^\beta}\right)^\omega
\end{split}
\end{equation}
\begin{equation}\label{eq8}
\begin{split}
\ln{L}_{LL3-WE3}&=n\ln\omega+n\ln\beta-n\beta\ln\lambda+(\beta-1)\sum_{i=1}^{n}\ln(x_{i}-\xi)+n\omega\ln\mu\\
&-\omega\sum_{i=1}^{n}\left(\frac{x_{i}-\xi}{\lambda}\right)^\beta
+(\omega-1)\sum_{i=1}^{n}\ln\left(1-(1+\delta)\exp\left[-\left(\frac{x_{i}-\xi}{\lambda}\right)^{\beta}\right]\right)\\
&-2\sum_{i=1}^{n}\ln\left[\left(1-(1+\delta)\exp\left[-\left(\frac{x_{i}-\xi}{\lambda}\right)^{\beta}\right]\right)^{\omega}+\left(\mu\exp\left[-\left(\frac{x_{i}-\xi}{\lambda}\right)^{\beta}\right]\right)^{\omega}\right]
\end{split}
\end{equation}
It should be noted that first-order partial derivatives of $\ln{L}$ in Eqs. \eqref{eq3}-\eqref{eq8} are ML estimations with respect to parameters of interest, so we can obtain the explicit solutions of log-likelihood equations for the parameters the usage the numerical methods such as conjugate gradients algorithm, Nelder–Mead, Brent–Dekker algorithm, quasi-Newton, and simulated annealing algorithm. The authors in this research to find the MLEs of the unknown parameters selected Nelder–Mead numerical method  as an optimization algorithm \cite{ref36}. Table \ref{tab3} indicates the estimated results of the parameters for WE3, LL3, LN3, GEV, WE3-LL3, and LL3-WE3 distributions via the MLE method.
\begin{table}[h!]
	\begin{center}\caption{\footnotesize{Computed parameters numerical values for the dataset (m/s) recorded from Tabriz station in 2018.}}
		\small

		\renewcommand{\arraystretch}{1.001}
	%	\resizebox{\textwidth}{!}{
			\begin{tabular}{cccccccc}

				\hline

				&Model&$\hat{\mu}$&$\hat{\omega}$&$\hat{\delta}$&$\hat{\lambda}$&$\hat{\beta}$&$\hat{\xi}$\\

				\hline

				Annual

				&WE3&1.6392&4.2325&-0.1573&-&-&-\\

				&LL3&0.2594&1.5289&-1.4711&-&-&-\\

				&LN3&0.4051&1.6539&-2.0398&-&-&-\\

				&GEV&0.0484&1.7539&2.5308&-&-&-\\

			    &WE3-LL3&0.9619&1.8805&0.7394&3.1318&0.7489&-2.3133\\
				&LL3-WE3&3.4232&1.5701&1.6756&6.1670&1.3805&-6.2797\\

				\\
				Winter

				&WE3&1.6941&3.0880&-0.1366&-&-&-\\

				&LL3&0.1529&1.6046&-2.5689&-&-&-\\

	            &LN3&0.3166&1.5231&-2.2014&-&-&-\\
				&GEV&0.0083&1.2494&1.9145&-&-&-\\

				&WE3-LL3&6.4659&2.2152&1.0062&0.2178&0.7331&-0.4352\\

				&LL3-WE3&3.4857&6.4302&6.1397&0.6950&0.3113&-8.6068\\

				\\
				Spring

				&WE3&1.6047&4.3832&-0.0706&-&-&-\\

				&LL3&0.2817&1.4683&-1.0607&-&-&-\\

				&LN3&0.4435&1.5950&-1.5831&-&-&-\\

				&GEV&0.0902&1.7636&2.6647&-&-&-\\

				&WE3-LL3&3.8217&1.2911&4.0028&7.5347&2.3723&-3.5170\\

				&LL3-WE3&1.9780&2.2705&0.6022&3.6228&0.7401&-1.7236\\

				\\
				Summer

				&WE3&2.1912&5.6465&-0.2672&-&-&-\\

				&LL3&0.1919&1.9458&-2.6297&-&-&-\\

				&LN3&0.2743&2.1288&-3.9934&-&-&-\\

				&GEV&-0.0736&2.0405&3.6808&-&-&-\\

				&WE3-LL3&2.3270&3.2969&2.5366&0.1830&0.4260&-2.1687\\

				&LL3-WE3&2.4518&3.9466&2.3182&3.6378&0.5017&-6.7806\\

				\\
				Autumn

				&WE3&1.5229&3.8098&-0.1539&-&-&-\\

		     	&LL3&0.2574&1.5297&-1.7658&-&-&-\\
			    &LN3&0.4019&1.6339&-2.2581&-&-&-\\
			    &GEV&0.0232&1.7329&2.2439&-&-&-\\
			    &WE3-LL3&8.2953&1.1520&5.9965&2.9900&2.0701&-7.1144\\
			    &LL3-WE3&3.1284&4.4782&1.1355&2.3734&0.4781&-4.3178\\
				\hline

		%\end{tabular}}\label{tab3}
\end{tabular}\label{tab3}
	\end{center}	
\end{table}
\subsection{Validation}
To evaluate different PDFs in terms of the modeling performance for a set of data, goodness-of-fit tests or various criteria have been applied to determine the best statistical models in studies concerning the analysis of data. Those most commonly applied are the KS test, $R^2$, RMSE, and $\chi^2$ test these criteria evaluate the deviation between the observed data using empirical distribution function (EDF) and the predicted data via theoretical CDF.\cite{ref18}. Also a descriptive of these indices exists in the following subsections.  
\subsubsection{Root mean square error}
The RMSE illustrates the actual deviation between predicted values of the model and observed cumulative probabilities\cite{ref34}. Also, the best distribution function model has the lowest value of RMSE. The RMSE is described below:
\begin{equation}
\mathrm{RMSE}=\sqrt{\frac{\sum_{j=1}^{n}(F_{j}-\hat{F}_{j})^2}{n}}
\end{equation}
where $\hat{F}_{j}$ is the predicted cumulative probability of the $j^{th}$ order statistics based on the proposed theoretical CDF and $F_{j}$ is the empirical cumulative probability of the $j^{th}$ order statistics with the formula:
\begin{equation}\label{eq2}
F_{j}=\frac{j-a}{n-2a+1} \qquad 0 \le a\le 1 
\end{equation}
where, $j=1,2,\dots,n$ is the rank for ascending ordered observations and $a$ denotes the plotting position constant\cite{ref37}. 
\subsubsection{Coefficient of determination}
The $R^2$ test is applied widely for hypothesis testing and goodness-of-fit comparisons because it quantifies the correlation between the predicted cumulative probabilities of a distribution model and the observed cumulative probabilities \cite{ref1}. The $R^2$ normally ranges from 0 to 1 and higher $R^2$ value indicates a better modeling performance of the theoretical distribution to the experimental data. The $R^2$ is defined as
\begin{equation}
R^2=1-\frac{\sum_{j=1}^{n}(F_{j}-\hat{F}_{j})^2}{\sum_{j=1}^{n}(F_{j}-\bar{F})^2}
\end{equation}
where $\bar{F}=\frac{\sum_{j=1}^{n}\hat{F_{j}}}{n}$, $\hat{F}_{j}$ is the predicted cumulative probability of the $j^{th}$ order statistics based on the proposed theoretical CDF and $F_{j}$ is the empirical cumulative probability of the $j^{th}$ order statistics with respect to Eq. \eqref{eq2}. 
\subsubsection{Kolmogorov-Smirnov (KS) test}
The KS test is the most popular goodness-of-fit test among other criteria that it evaluates the maximum error between the CDF of the fitted data and the EDF of  the observed data \cite{ref34}. The KS is expressed by the below equation: 
\begin{equation}
D_{n}=\underset{1\le j\le n}{\sup}|F_{j}-\hat{F}_{j}|
\end{equation} 
where $F_{j}$ is the empirical cumulative probability of the $j^{th}$ order statistics with respect to Eq. \eqref{eq2} and $\hat{F}_{j}$ is the predicted cumulative probability of the $j^{th}$ order statistics based on the proposed theoretical CDF. The lowest value of KS shows that the considered distribution is the best model.
\subsubsection{Chi-square ($\chi^2$) test}
The Chi-square test is applied to calculate whether the predicted cumulative probability of theoretical CDF differs from the observed cumulative probability. It should be noted a lesser value of this test statistic illustrates the better modeling performance of the PDF. \cite{ref19}. The Chi-Square test statistic is described as follows:
\begin{equation}
\chi^2=\sum_{j=1}^{n}\frac{(F_{j}-\hat{F_{j}})^2}{\hat{F_{j}}}
\end{equation}
where $F_{j}$ is the empirical cumulative probability of the $j^{th}$ order statistics with respect to Eq. \eqref{eq2} and $\hat{F}_{j}$ is the predicted cumulative probability of the $j^{th}$ order statistics based on the proposed theoretical CDF.
\begin{table}[h!]
	\begin{center}\caption{\footnotesize{Assessment of statistical errors with ranked PDFs for the dataset recorded from Tabriz station in 2018.}}
		\footnotesize

		\renewcommand{\arraystretch}{0.97}

	%	\resizebox{\textwidth}{!}{
			\begin{tabular}{ccccccc}

				\hline

				&Model&KS&$R^2$&RMSE&$\chi^2$&Rank\\

				\hline

				Annual

				&WE3&0.1354&0.9706&0.0495&23.9369&6\\

				&LL3&0.1073&0.9780&0.0428&18.8253&2\\

				&LN3&0.1124&0.9771&0.0436&19.5499&3\\

				&GEV&0.1213&0.9755&0.0452&21.6542&4\\

			&WE3-LL3&0.1321&0.9719&0.0484&23.3270&5\\

			&LL3-WE3&\textbf{0.0992}&\textbf{0.9793}&\textbf{0.0415}&\textbf{18.0069}&\textbf{1}\\

				\\
				Winter

				&WE3&0.2107&0.9070&0.0879&13.6159&6\\

			&LL3&0.1458&0.9484&0.0655&9.5749&2\\

			&LN3&0.1793&0.9310&0.0757&11.1246&3\\

			&GEV&0.1802&0.9303&0.0761&11.6164&4\\

			&WE3-LL3&0.2080&0.9092&0.0868&13.2413&5\\

			&LL3-WE3&\textbf{0.1455}&\textbf{0.9487}&\textbf{0.0653}&\textbf{9.4984}&\textbf{1}\\

				\\
				Spring

				&WE3&0.1376&0.9678&0.0517&6.4234&5\\

			&LL3&0.1047&0.9798&0.0409&4.3847&2\\

			&LN3&0.1110&0.9781&0.0426&4.6806&4\\

			&GEV&0.1085&0.9796&0.0411&4.4638&3\\

			&WE3-LL3&0.1475&0.9602&0.0575&7.5979&6\\

			&LL3-WE3&\textbf{0.0981}&\textbf{0.9814}&\textbf{0.0394}&\textbf{4.1034}&\textbf{1}\\

				\\
				Summer

				&WE3&0.0713&0.9882&0.0313&2.8987&2\\

			&LL3&0.0833&0.9859&0.0342&3.1690&6\\

			&LN3&0.0763&0.9884&0.0311&2.7490&3\\

			&GEV&0.0788&0.9879&0.0318&2.8311&4\\

			&WE3-LL3&\textbf{0.0709}&\textbf{0.9889}&\textbf{0.0303}&\textbf{2.7284}&\textbf{1}\\

			&LL3-WE3&0.0791&0.9867&0.0332&3.0047&5\\

				\\
				Autumn

				&WE3&0.1371&0.9753&0.0453&6.0072&5\\

			&LL3&0.1048&0.9792&0.0416&4.4866&2\\

			&LN3&0.1082&0.9789&0.0415&4.6498&3\\

			&GEV&0.1147&0.9787&0.0421&4.6711&4\\

			&WE3-LL3&0.1512&0.9713&0.0488&7.3538&6\\

			&LL3-WE3&\textbf{0.1021}&\textbf{0.9800}&\textbf{0.0408}&\textbf{4.4397}&\textbf{1}\\

				\hline

		\end{tabular}\label{tab4}
	\end{center}	
\end{table}
\begin{figure}[h!]
	\begin{center}
		\subfigure[Winter]{
			\includegraphics[height=6.5cm,width=6.5cm]{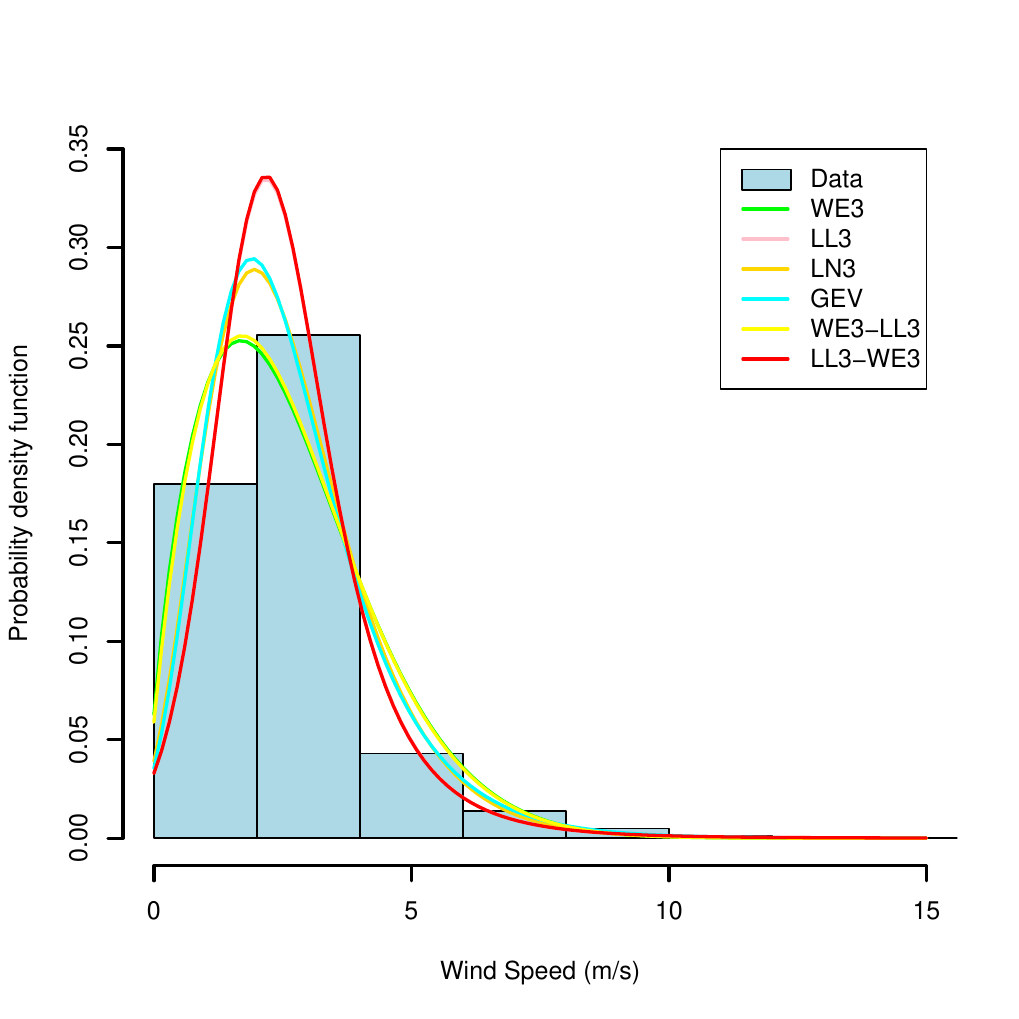}}
		\hspace{0.5cm}
		\subfigure[Spring]{
			\includegraphics[height=6.5cm,width=6.5cm]{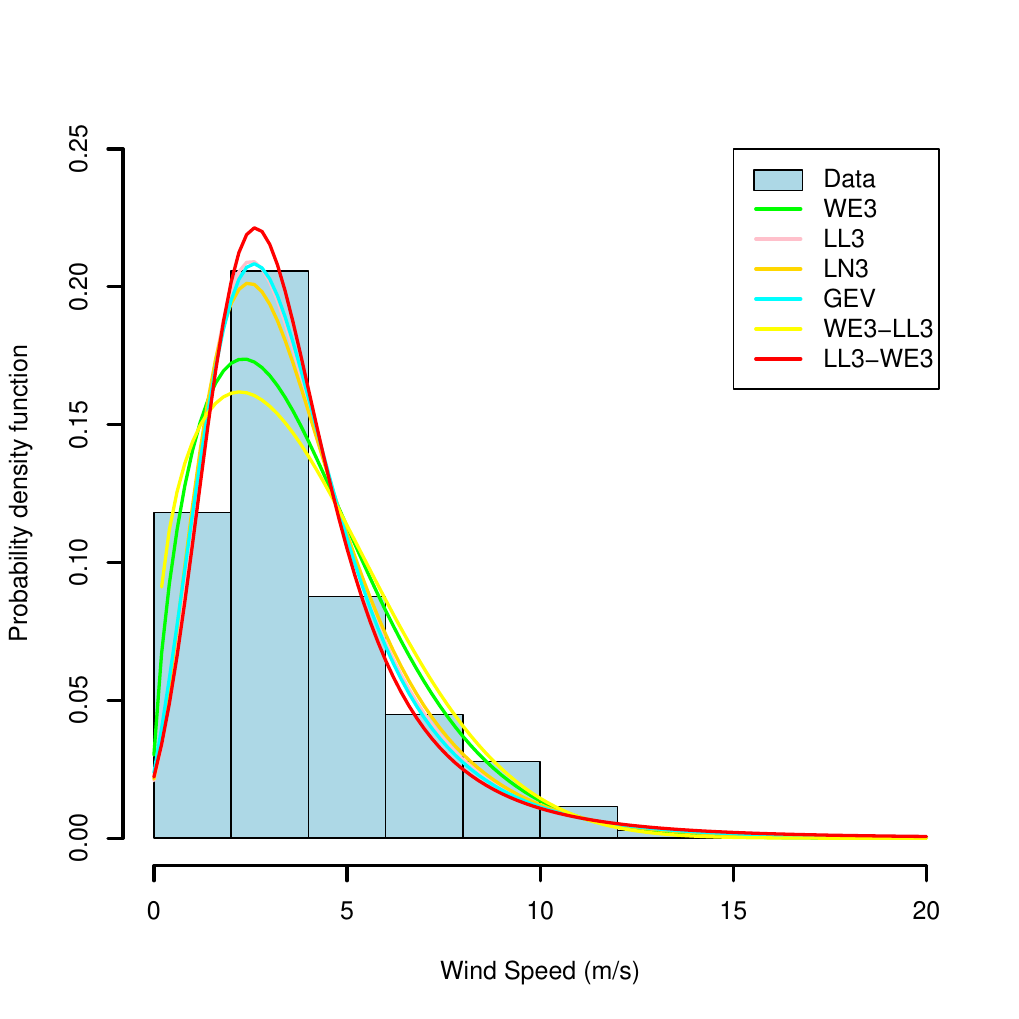}}
		\hspace{0.5cm}
		\subfigure[Summer]{
			\includegraphics[height=6.5cm,width=6.5cm]{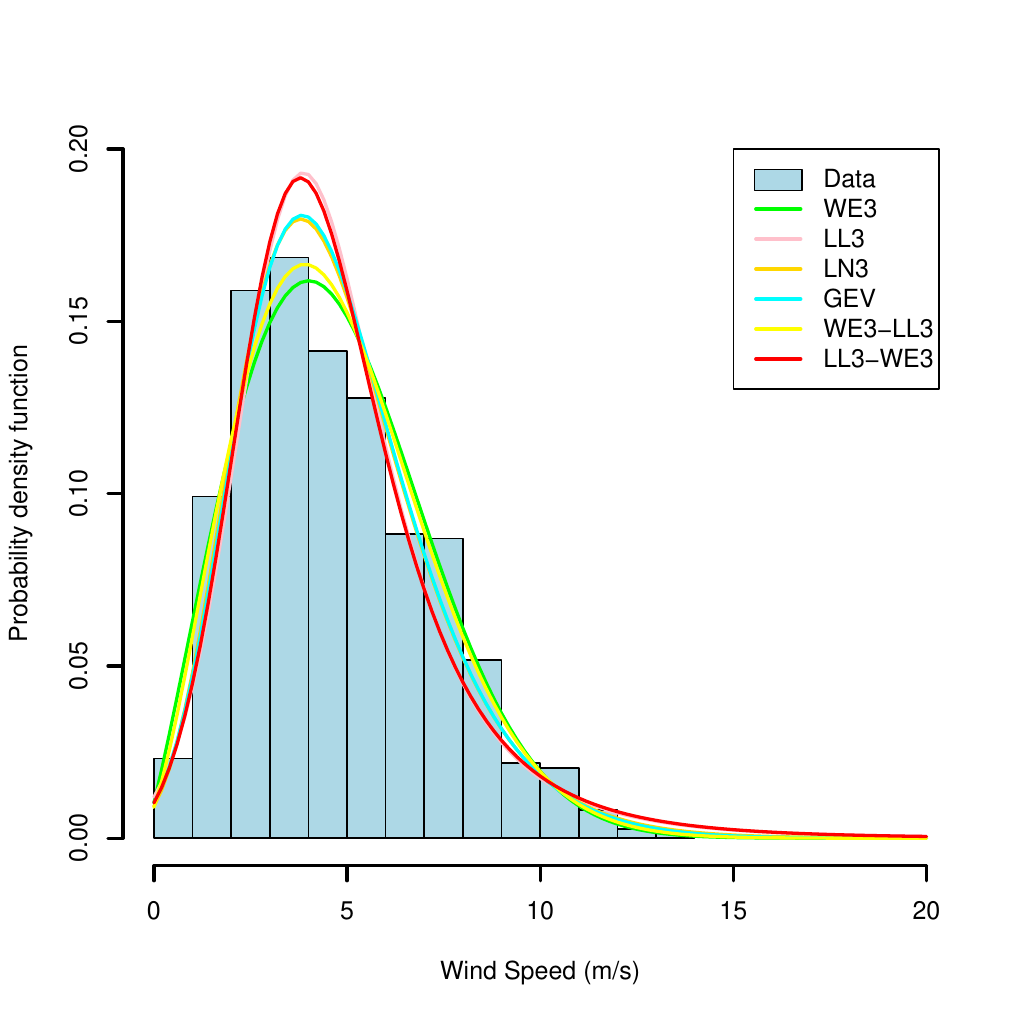}}
		\hspace{0.5cm}
		\subfigure[Autumn]{
			\includegraphics[height=6.5cm,width=6.5cm]{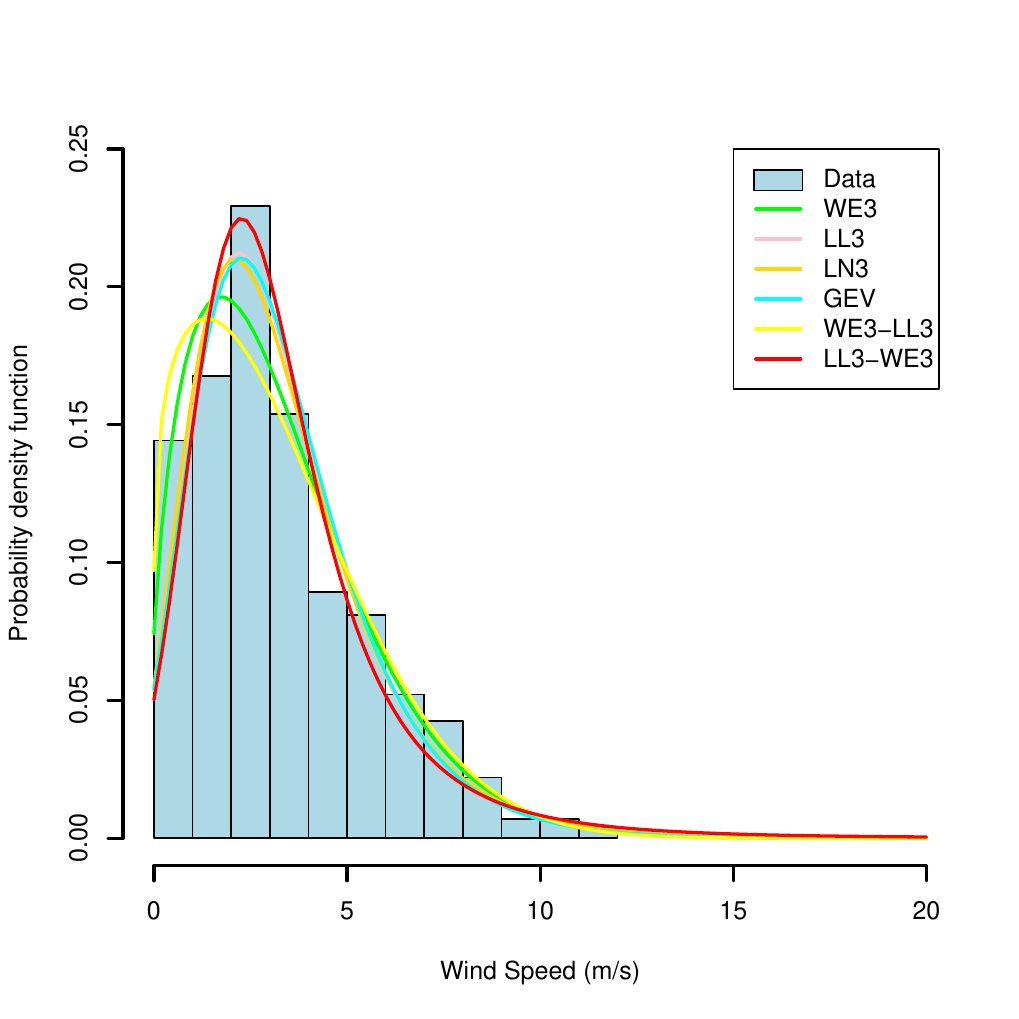}}
		\hspace{0.5cm}
		\subfigure[Annual]{
			\includegraphics[height=6.5cm,width=6.5cm]{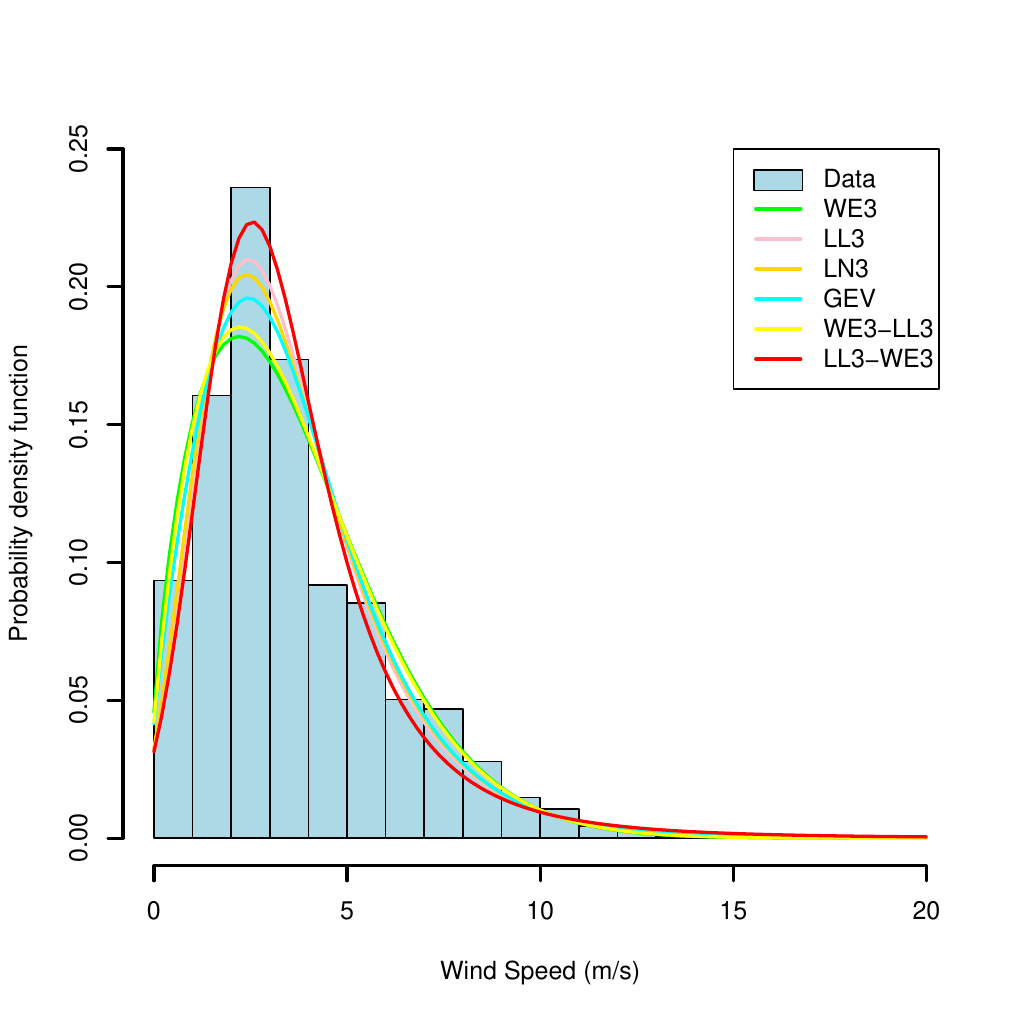}}	
	\end{center}\caption{\footnotesize{Graphical Comparison of the fitted PDFs and histograms of the data recorded from Tabriz station in 2018.}}
	\label{fig3}
\end{figure}
\section{Results and discussion}\label{sec5}
In this paper, we evaluate the fitting performance of considered different distributions in terms of KS, $R^2$, RMSE, and $\chi^2$ criteria for the wind speed data set (m/s) of Tabriz station in Iran. Table \ref{tab4} illustrates values of goodness-of-fit tests for each fitted distribution model. The last column of Table \ref{tab4} presents the ranked distribution functions of considered models. The LL3-WE3 distribution provides a better fitting than the other PDFs to evaluate the annual and seasonal wind speed data except summer season, hence LL3-WE3 distribution has the biggest $R^2$ value and the smallest numerical values of $\chi^2$, RMSE, and KS. Besides, the WE3-LL3 model is introduced as the most suitable distribution function based on all statistical parameters to analyze wind speed data in summer.\\
The analyzed results of statistical indicators are also in the matching with the histogram and curves of fitted density functions shown in Fig. \ref{fig3}. As mentioned earlier, WE3-LL3 distribution demonstrates the appropriate modeling performance for the wind speed data reported in the summer, see Fig. \ref{fig3}(c). It can also be observed from Fig. \ref{fig3}(a), (b), (d), and (e) that LL3-WE3 distribution is the best model among the proposed distribution functions to model wind speed data in winter, spring, autumn, and annual in Tabriz station.\\
In addition to goodness-of-fit tests, the wind power density error (PDE) criterion can be applied to show the capability of distribution in evaluating wind power. The PDE is measured using reference mean wind power density and mean wind power density based on the distributional model, also it should be emphasized that the smallest value of PDE provides a better fitting. The PDE is formulated as
\begin{equation}
PDE=\left|\frac{P_{ref}-P_{D}}{P_{ref}}\right|\times100\end{equation}
where ${P_{D}}$ is the mean wind power density with respect to the distributional model and it is given as follows:
\begin{equation}
P_{D}=\frac{1}{2}\rho A \int_{0}^{\infty}x^{3}f_{D}(x)\mathrm{d}x 
\end{equation} 
also, $P_{ref}$ is the reference mean wind power density on real wind speed data and it is given by
\begin{equation}
P_{ref}=\frac{1}{2}\rho A \frac{1}{n}\sum_{i=1}^{n}{x_{i}}^3
\end{equation}
Here, $A$ is  wind turbine blade sweep area ($\mathrm{m^2}$) and $\rho$ is air density ($\mathrm{kg/m^3}$). 
\begin{table}[h!]
	\begin{center}\caption{\footnotesize{Results of PDE criterion and Wind power density for the dataset recorded from Tabriz station in 2018.}}
		\footnotesize

		\renewcommand{\arraystretch}{1}

		%\resizebox{\textwidth}{!}{

			\begin{tabular}{cccccc}

				\hline

				&Annual&Winter&Spring&Summer&Autumn\\

				\hline

				P$_{REF}$&125.26&51.76&154.55&196.72&96.01\\

				P$_{WE3}$&121.69&44.05&145.31&195.57&99.95\\

				P$_{LL3}$&250.94&40.83&374.15&275.44&223.81\\

				P$_{LN3}$&129.04&41.14&167.08&203.07&106.05\\

				P$_{GEV}$&131.29&40.25&166.14&197.04&102.77\\

				P$_{WE3-LL3}$&122.17&43.76&146.67&196.52&99.56\\

				P$_{LL3-WE3}$&124.06&44.42&155.89&264.54&97.28\\

				\\
				PDE$_{WE3}$&2.85&14.89&5.98&0.58&4.10\\

				PDE$_{LL3}$&100.33&21.12&142.09&40.02&133.12\\

				PDE$_{LN3}$&3.01&20.52&8.11&3.23&10.46\\

				PDE$_{GEV}$&4.81&22.24&7.50&0.16&7.05\\

				PDE$_{WE3-LL3}$&2.47&15.46&5.10&\textbf{0.10}&3.70\\

				PDE$_{LL3-WE3}$&\textbf{0.96}&\textbf{14.19}&\textbf{0.87}&34.48&\textbf{1.33}\\

				\hline
		\end{tabular}\label{tab5}
	\end{center}
\end{table}\\
Table \ref{tab5} shows the estimated experimental and analytical values of wind power density and also PDE values for the chosen distributions. It should be noted that PDE is a model evaluation criterion and its value depends on the applied parameters estimation methods and the chosen distribution models. However, the accurate model with the ability of the high performance shows the lowest error in wind power among the weak performances. Similar to statistical parameters, LL3-WE3 distribution provides the most favorable fit to the PDE criterion to analyze the annual and seasonal wind speed data except summer, because LL3-WE3 has the lowest PDE value than the other models as well as WE3-LL3 distribution function has a good fitting ability for modeling the wind data in the summer season.\\
Generally, in this section, we concluded  that LL3-WE3 and WE3-LL3 distributions are more flexible than other PDFs of terms in goodness-of-fit tests and PDE criteria for seasonal and annual wind speed data recorded of Tabriz location.
\section{Conclusion}\label{sec6} 
In the present study, a comparison of the performance of the wind speed distribution functions was evaluated to show the most successful models in the seasonal and annual wind speed data analysis from the Tabriz meteorological station in Iran. The authors suggested a new approach to generate and develop the new probability density functions for wind speeds through the T-X family of continuous distributions that was proposed for the first time in this research. The MLE method is presented as an effective technique to estimate parameters of interest of the considered distributions by using the Nelder–Mead optimization method. The results obtained via the model selection criteria namely KS, RMSE, $\chi^2$, and $R^2$ illustrate that LL3-WE3 distribution is the most suitable distribution function for modeling wind speed data except summer. Also, WE3-LL3 distribution provides the best fit to estimate wind data in the summer season. Similarly, the values reported by the PDE indicator express that LL3-WE3 and WE3-LL3 models are preferred over the other PDFs to minimize the error of wind power density for the wind speed data measured in the considered period.\\
Overall, the results of our analyses indicate that the generated distributions (LL3-WE3 and WE3-LL3) of the T-X family have the higher flexibility than the other models to describe wind speed characteristics. Therefore, they can be used as a helpful alternative to the 2-parameter Weibull in the wind energy literature. Finally, we recommend employing every one of these two distribution functions for the estimation of wind energy potential in the variant wind speed regimes in other countries.
	

\begin{thebibliography}{99}
\bibitem{ref1}
Akgul FG, Senoglu B, Arslan T. An alternative distribution to Weibull for modeling
the wind speed data: inverse Weibull distribution. Energy Convers Manage
2016;114:234–40.
\bibitem{ref2}
Wang J, Quin S, Jin S, Wu J. Estimation methods review and analysis of offshore extreme wind speeds and wind energy resources. Renew Sustain Energy Rev 2015;42:26–42.
\bibitem{ref3}
Masseran N. Evaluating wind power density models and their statistical properties. Energy 2015;84:533–41.
\bibitem{ref4}
Dai K, Bergot A, Liang C, Xiang WN, Huang Z. Environmental issues associated with wind energy–a review. Renew Energy 2015;75:911–21.
\bibitem{ref5}
Sukru A, Senoglu B, Arslan T. Generalized Lindley and Power Lindley distributions for modeling the wind speed data . Energy Convers Manage 2017;152:300–311.
\bibitem{ref6}
Huang J, McElroy MB. A 32-year perspective on the origin of wind energy in warming climate. Renew Energy 2015;77:482–92.
\bibitem{ref7}
Mohammadi K, Alavi O, Mostafaeipour A, Goudarzi N, Jalilvand M. Assessing different parameters estimation methods of Weibull distribution to compute wind power density. Energy Convers Manage 2016;108:322–35.
\bibitem{ref8}
Xu X, Yan Z, Xu S. Estimating wind speed probability distribution by diffusionbased kernel density method. Electr Power Sys Res 2015;121:28–37.
\bibitem{ref9}
Kantar YM, Usta I. Analysis of the upper-truncated Weibull distribution for wind speed. Energy Convers Manage 2015;96:81–8.
\bibitem{ref10}
Rocha PAC, Sousa RC, Andrade CF, Silva MEV. Comparison of seven numerical methods for determining Weibull parameters for wind energy generation in the northeast region of Brazil. Appl Energy 2012;89:395–400.
\bibitem{ref11}
Usta I, Kantar YM. Analysis of some flexible families of distributions for estimation of wind speed distributions. Appl Energy 2012;89:355–67.
\bibitem{ref12}
Arslan T, Bulut YM, Yavuz AA. Comparative study of numerical methods for determining Weibull parameters for wind energy potential. Renew Sustain Energy Rev 2014;40:820–5.
\bibitem{ref13}
Khahro SF, Tabbassum K, Soomro AM, Dong L, Liao X. Evaluation of wind power production prospective and Weibull parameter estimation methods for Babaurband, Sindh Pakistan. Energy Convers Manage 2014;78:956–67.
\bibitem{ref14}
Akdag SA, Guler O. A novel energy pattern factor method for wind speed distribution parameter estimation. Energy Convers Manage 2015;106:1124–33.
\bibitem{ref15}
Shu ZR, Li QS, Chan PW. Statistical analysis of wind characteristics and wind energy potential in Hong Kong. Energy Convers Manage 2015;101:644–57.
\bibitem{ref16}
Bilir L, Imir M, Devrim Y, Albostan A. An investigation on wind energy potential and small scale wind turbine performance at Incek region – Ankara, Turkey. Energy Convers Manage 2015;103:910–23.
\bibitem{ref17}
Kaplan YA. Overview of wind energy in the world and assessment of current wind energy policies in Turkey. Renew Sustain Energy Rev 2015;43:562–8.
\bibitem{ref18}
Masseran N. Integrated approach for the determination of an accurate wind-speed distribution model. Energy Convers Manage 2018;173:56-64.
\bibitem{ref19}
Ouarda T.B.M.J, Charron C, Shin J-Y, Marpu P.R, Al-Mandoos A.H, Al-Tamimi M.H, Ghedira H, Al Hosary T.N. Probability distributions of wind speed in the UAE. Energy Convers Manage 2015;93:414-434.
\bibitem{ref20}
Aries N, Boudia SM, Ounis H. Deep assessment of wind speed distribution models: a case study of four sites in Algeria. Energy Convers Manage 2018;155:78–90.
\bibitem{ref21}
Morgan EC, Lackner M, Richard MV, Baise LG. Probability distributions for offshore wind speeds. Energy Convers Manage 2011;52:15–26.
\bibitem{ref22}
Stewart DA, Essenwanger OM. Frequency distribution of wind speed near the surface. J Appl Meteorol 1978;17(11):1633–42.
\bibitem{ref23}
Tuller SE, Brett AC. The goodness of fit of the Weilbull and Rayleigh distribution to the distributions of observed wind speeds in a topographically diverse area. J Climatol 1985;5:74–94.
\bibitem{ref24}
Kidmo D.K, Danwe R, Doka S.Y, Djongyang N. Statistical analysis of wind speed distribution based on six Weibull Methods for wind power evaluation in Garoua, Cameroon. Revue des Energies Renouvelables 2015;18(1):105-125.
\bibitem{ref25}
Jung C, Schindler D. Global comparison of the goodness-of-fit of wind speed distributions. Energy Convers Manage 2017;133:216–34.
\bibitem{ref26}
Abdulkarim A, Abdelkader S.M, Morrow D.J, Falade A.J, Adediran Y.A. Statistical analysis of wind speed for electrical power generation in some selected sites in northern Nigeria. Nigerian Journal of Technology 2017;36(4):1249-1257.
\bibitem{ref27}
Lee BH, Ahn DJ, Kim GH, Ha YC. An estimation of the extreme wind speed using the Korea wind map. Renew Energy 2012;42:4–10.
\bibitem{ref28}
Kantar YM, Usta I, Arik A, Yenilmez I. Wind speed analysis using the Extended Generalized Lindley Distribution. Renew Energy 2018;118:1024–30.
\bibitem{ref29}
Brano VL, Orioli A, Ciulla G, Culotta S. Quality of wind speed fitting distributions for the urban area of Palermo, Italy. Renew Energy 2011;36:1026–39.
\bibitem{ref30}
Masseran N, Razali AM, Ibrahim K, Zaharim A, Sopian K. Application of the single imputation method to estimate missing wind speed data in Malaysia. Res J Appl Sci, Eng Technol 2013;6:1780–4.
\bibitem{ref31}
Alzaatreh A, Lee C, Famoye F. A new method for generating families of continuous distributions. Metron 2013;71(1):63-79.
\bibitem{ref32}
Aljarrah MA, Lee C, Famoye F. On generating TX family of distributions using quantile functions. Journal of Statistical Distributions and Applications 2014;1(1):2.
\bibitem{ref33}
Singh V.P, Guo H, Yu F.X. Parameter estimation for 3-parameter log-logistic distribution (LLD3) by Pome. stochastic hydrology and hydraulics 1993;7:163-177.
\bibitem{ref34}
Moyazzem Hossain MD. Probability Modeling of Monthly Maximum Sustained Wind Speed in Bangladesh. Statistics optimization and information computing 2019;7:75-84.
\bibitem{ref35}
Ying A and Pandey M.D. The r largest order statistics model for extreme wind speed estimation. J Wind Eng and Aerodyn 2007;95(3):165-182.
\bibitem{ref36}
Koca M.B, Kiliç M.B and Şahin Y. Assessing wind energy potential using finite mixture distributions. Turkish Journal of Electrical Engineering and Computer Sciences 2019;27:2276-2294.
\bibitem{ref37}
Cunnane C. Unbiased plotting positions – a review. J Hydrol 1978;37(3–4):205–22.

\end{thebibliography}
\end{document}